\documentclass[useAMS,usenatbib]{mn2e}

\title[Near-IR and optical studies of nova V496 Sct]{ V496 Scuti: An Fe II nova with dust shell accompanied by CO emission.}

\author[ Raj et al.]{ Ashish Raj$^{1}$, N.M. Ashok$^{1}$, D.P.K. Banerjee$^{1}$,U. Munari$^{2}$, P. Valisa$^{3}$, S. Dallaporta$^{3}$ \\
$^{1}$Astronomy and Astrophysics Division, Physical Research Laboratory, Navrangpura, Ahmedabad - 380009, Gujarat, India\\
$^{2}$INAF Astronomical Observatory of Padova, 36012 Asiago (VI), Italy\\
$^{3}$ANS Collaboration, c/o Osservatorio Astronomico, via dell'Osservatorio 8, 36012 Asiago (VI), Italy}

\usepackage[dvips]{graphicx}

\begin{document}

\date{Accepted  Received }

\pagerange{\pageref{firstpage}--\pageref{lastpage}} \pubyear{2010}

\maketitle

\label{firstpage}

\begin{abstract}
We present near-infrared and optical observations of the nova V496 Scuti 2009 covering various phases - pre-maximum, early decline and nebular - during the first 10 months of its discovery followed by limited observations in early part of 2011 April. The spectra follow the evolution of the nova when the lines had strong P Cygni profiles to a phase dominated by prominent emission lines. The notable feature of the near-IR spectra in the early decline phase is the rare presence of first overtone bands of carbon monoxide in emission. Later about 150 days after the peak brightness the IR spectra show clear dust formation in the expanding ejecta. Dust formation in V496 Sct is consistent with the presence of lines of elements with  low ionization potentials like Na and Mg in the early spectra and the detection of CO bands in emission. The light curve shows a slow rise to the maximum and a slow decline indicating a prolonged mass loss. This is corroborated by the strengthening of P Cygni profiles during the first 30 days. In the spectra taken close to the optical maximum brightness, the broad and single absorption component seen at the time of discovery is replaced by two sharper components. During the early decline phase two sharp dips that show increasing outflow velocities are seen in the P Cygni absorption components of Fe II and H I lines. The spectra in 2010 March showed the onset of the nebular phase. Several emission lines display saddle-like profiles during the nebular phase. In the nebular stage the observed fluxes of [O III] and H$\beta$ lines are used to estimate the electron number densities and the mass of the ejecta. The optical spectra show that the nova evolved in the P$_{fe}$A$_{o}$ spectral sequence. The physical conditions in the ejecta are estimated. The absolute magnitude and the distance to the nova are estimated to be $M_V = -7.0$ $\pm$ 0.2 and $d = 2.9$ $\pm$ 0.3 kpc respectively.
\end{abstract}


\section{Introduction}
Nova Scuti 2009 (V496 Sct) was discovered by Nishimura on 2009 November 8.370 UT at $V = 8.8$ (Nakano 2009). The low resolution spectra obtained soon after its discovery in the period 2009 November 9.73 UT to 10.08 UT showed prominent H$\alpha$ and H$\beta$ emission lines with P Cygni components, along with the strong Fe II multiplets and O I lines indicating that V496 Scuti is an Fe II class nova near maximum light (Teyssier 2009, Munari et al. 2009a, Balam \& Sarty 2009). The typical FWHM of the P Cygni components ranged from 700 to 950 km s$^{-1}$ with the absorption component blue shifted by 700 km s$^{-1}$. The follow-up observations by Munari et al. (2009b) showed a post-discovery brightening for about 10 days before the onset of fading with maximum brightness $V_{max} = 7.07$ around 2009 November 18.716 UT. V496 Sct was observed  in the infrared by Rudy et al. (2009) using Near Infrared Imaging Spectrograph on the 3m Shane reflector at Lick Observatory on 2009 November 27.08 UT and revealed strong first overtone CO emission - an extremely short lived feature that is seen in only a few novae. They also found several prominent C I emission lines with the strongest line accompanied by P Cygni type absorption component. As the novae that display first overtone CO in emission and strong C I emission in early phases form dust, Rudy et al. (2009) predicted that dust formation in V496 Sct is almost certain. Following this interesting prediction an observational campaign of V496 Sct was initiated at Mt. Abu IR Observatory and the first result by Raj, Ashok \& Banerjee (2009) showed the continuation of CO emission during the period 2009 December 3.55 UT to 8.55 UT. Subsequent observations by Russell et al. (2010) after V496 Sct came out of the solar conjunction showed dust formation on 2010 February 10. The CO emission seen in 2009 November was absent.

An inspection by Guido \& Sostero (2009) of the Digitized Sky Survey (DSS) plate (limiting red magnitude about 20) obtained on 1996 August 13  did not reveal any clear and unambiguous object at the position of V496 Sct. The limiting red magnitude of 20 for the DSS plate makes V496 Sct one of the large amplitude ($\Delta$R $\ge$ 13.5) novae.

The nova V496 Sct was studied at Mt. Abu IR Observatory of Physical Research Laboratory in India, at Asiago Observatory operated by the University of Padova and INAF Astronomical Observatory of Padova and Schiaparelli Observatory in Italy. In this paper we present spectral evolution during the pre-maximum rise, early decline and the nebular phase.

\section{Observations}
\subsection{Infrared observations}
Near-IR observations were obtained using the 1.2m telescope of Mt.Abu Infrared Observatory from 2009 November 19 to 2011 April 23. The availability of V496 Sct for short duration during 2009 December resulted in restricted photometric coverage. The log of the spectroscopic and photometric observations are given in Table 1. The spectra were obtained at a resolution of $\sim$ 1000 using a Near-Infrared Imager/Spectrometer
with a 256$\times$256 HgCdTe NICMOS3 array. In each of the $JHK$ bands a set of spectra was taken with the nova
off-set to two different positions along the slit ( slit width 1 arc second). The wavelength calibration was done using
the OH sky lines that register with the stellar spectra. The spectra of the comparison star SAO 144150 (spectral type B9.5 III) and SAO 142612 (spectral type B9) were taken at similar airmass as that of V496 Sct to ensure that the ratioing process (nova spectrum divided by the standard
star spectrum) removes the telluric features reliably. To avoid artificially generated emission lines in the
ratioed spectrum, the  H\,{\sc i} absorption lines in the spectra of standard star were removed by interpolation
before ratioing. The ratioed spectra were then multiplied by a blackbody curve corresponding to the standard star's
effective temperature to yield the final spectra.

Photometry in the $JHK$ bands was done  in clear sky conditions using the NICMOS3 array in the imaging mode.
Several frames,  in 4 dithered positions, offset by $\sim$ 30 arcsec were obtained in all the bands. The sky
frames, which are subtracted from the nova frames, were generated by median combining the dithered frames.
The star SAO 142612 having 2MASS $JHK$ magnitudes 6.55, 6.51 and 6.44 respectively and located close to the nova was used for photometric calibration. The data is reduced and analyzed using the
IRAF package.
  \begin{figure}
\includegraphics[width=3.0in,height=5.5in,clip]{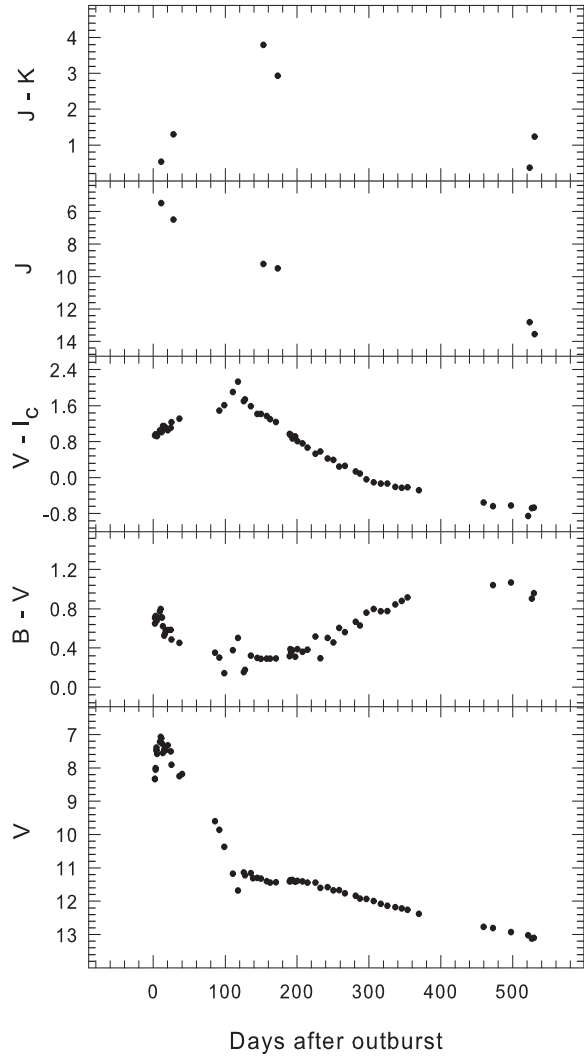}
\caption[]{ The light curve of V496 Sct from Asiago and Mt. Abu data.}
 \label{fig1}
 \end{figure}


\begin{table*}
\centering
\caption{Log of the Mt. Abu near-infrared observations of V496~Sct. The
date of optical maximum is taken as 2009 November 18.716 UT.}
\begin{tabular}{ccccccccccc}
\hline
Date of      &Days since    &\multicolumn{3}{|c|}{Integration time (s)} &\multicolumn{3}{|c|}{Integration time (s)} &\multicolumn{3}{|c|}{Nova Magnitude} \\
Observation  &optical maximum             &J-band      &H-band   &K-band  &J-band      &H-band   &K-band   &J-band      &H-band   &K-band  \\
\hline
   &   &\multicolumn{3}{|c|}{Spectroscopic Observations}   &\multicolumn{4}{|l|}{Photometric Observations}  \\
\hline
2009 Nov. 19.57	&0.854    &60  &--   &--   &10  &25  &50  &5.49$\pm$0.02 &5.20$\pm$0.02  &4.96$\pm$0.10 \\
2009 Dec. 03.58	&14.864    &--  &--   &60  &--  &--  &--  &-- &--  &-- \\
2009 Dec. 05.56	&16.844    &60  &60   &120  &--  &--  &--  &-- &--  &-- \\
2009 Dec. 06.56	&17.844   &90  &60   &120   &5  &10  &10  &6.50$\pm$0.16 &5.82$\pm$0.21  &5.21$\pm$0.14 \\
2009 Dec. 07.54	&18.824   &--  &--   &120   &--  &--  &25 &-- &--  &5.67$\pm$0.20 \\
2009 Dec. 08.54	&19.824   &90  &70   &120   &--  &--  &25 &-- &--  &5.45$\pm$0.20\\
2009 Dec. 09.22	&20.504   &-- &90  &120  &--  &--  &25  &-- &--  &5.27$\pm$0.07\\
2010 Apr. 10.97	&143.254   &-- &--  &--  &250  &275  &50  &9.24$\pm$0.15 &7.74$\pm$0.15  &5.45$\pm$0.20\\
2010 Apr. 11.97	&144.254   &120 &120  &80  &--  &--  &--  &-- &-- &-- \\
2010 Apr. 21.94	&154.224   &300 &200  &200  &--  &--  &-- &-- &--  &-- \\
2010 Apr. 22.94	&155.224   &120 &80  &60  &--  &--  &-- &-- &--  &-- \\
2010 Apr. 23.93	&156.214   &120  &80 &60  &-- &-- &-- &-- &-- &--\\
2010 Apr. 29.92	&162.204   &150  &100 &100  &--  &--  &-- &-- &--  &-- \\
2010 Apr. 30.92	&163.204   &300  &-- &--  &625 &550 &50 &9.50$\pm$0.09 &7.99$\pm$0.09 &6.57$\pm$0.10\\
2011 Apr. 15.95	&513.234   &--  &-- &--  &1000 &500 &105 &12.82$\pm$0.15 &13.03$\pm$0.15 &12.46$\pm$0.20\\
2011 Apr. 22.96	&520.244   &--  &-- &--  &1000 &750 &105 &13.55$\pm$0.12 &13.28$\pm$0.10 &12.32$\pm$0.12\\
\hline
\end{tabular}
\label{table1}
\end{table*}

\subsection{Optical spectroscopy}

Spectroscopic observations of V496 Sct were obtained with the 0.6m
telescope of the Schiaparelli observatory in Varese, equipped with a multi
mode spectrograph (Echelle + single dispersion modes) and various reflection
gratings, as part of the ANS Collaboration monitoring of nova outbursts
(Munari et al.  2011).  A journal of the spectroscopic observations is
provided in Table~3, where the time is counted from the $V$ band maximum.
The resolving power of the Echelle spectra is $\sim$20,000.  The spectra
were exposed with a 2 arcsec wide slit, oriented along the instantaneous
parallactic angle.  All spectra (including Echelle ones) were calibrated in
absolute fluxes by observations of several spectrophotometric standards each
night, at similar airmasses and both immediately before and after the
exposure on the nova.  Their zero-points were then checked against
simultaneous BVRI photometry by integrating the band transmission profiles
on the fluxed spectra, with the differences almost never exceeding 0.1 mag.

\subsection{Optical photometry}

Optical photometry of V496 Sct was obtained with ANS Collaboration
telescope number R030 located in Cembra (Trento, Italy). A detailed
description of ANS Collaboration instruments, operation modes and results on
the monitoring of novae is provided by Munari et al.  (2011) and Munari \&
Moretti (2011).  Telescope R030 is a 0.30-m Meade RCX-400 f/8
Schmidt-Cassegrain telescope, equipped with a SBIG ST-9 CCD camera,
512$\times$512 array, 20 $\mu$m pixels $\equiv$ 1.72$^{\prime\prime}$/pix,
providing a field of view of 13$^\prime$$\times$13$^\prime$.  The $B$ filter
is from Omega and the $U$$V$$R_{\rm C}$$I_{\rm C}$ filters from Custom
Scientific. The $B$$V$$R_{\rm C}$$I_{\rm C}$ photometry of Nova Sct 2009 is
presented in Table~2. The median values for the total error budget are
$\sigma$(V)=0.008, $\sigma$(B-V)=0.014, $\sigma$(V-R)=0.012,
$\sigma$(V-I)=0.027, which include both the Poissonian components and
the uncertainty in the transformation from the local to the standard
Landolt (1992) system.

\begin{table}
\caption{Log of the optical photometric observations of V496 Sct. The date of optical maximum
is taken to be 2009 November 18.716 UT.}

\begin{tabular}{lccccccc}
\hline
Date & Days since         & &             &           &          &       \\
     & optical        & &             &           &          &        \\
(UT) &  maximum      &\emph{V} & \emph{B-V}     &  \emph{V-Rc} &\emph{V-Ic} & \\
\hline
2009        &       &      &     &      &      & \\
Nov. 10.692	& -8.024 &8.341 &0.647&0.471 &0.933 &	\\
Nov. 10.710	& -8.006 &8.329 &0.707& & &	\\
Nov. 11.693	& -7.023 &8.057 &0.725&0.506 &0.961 &	\\
Nov. 11.705	& -7.011 &8.009 &0.709& &  &	\\
Nov. 12.698	& -6.018 &7.393 &0.712& & &	\\
Nov. 12.700	& -6.016 &7.463 &0.680&0.479 &0.935 &	\\
Nov. 12.736	& -5.980 &7.487 &0.672&0.469 &0.913 &	\\
Nov. 13.732	& -4.984 &7.585 &0.711&0.457 &0.914 &	\\
Nov. 17.706	& -1.010 &7.219 &0.772&0.508 &1.051 &	\\
Nov. 18.716	& 0.000 &7.070 &0.797&0.526 &1.037 &	\\
Nov. 19.710	& 0.994 &7.115 &0.713&0.472 &1.000 &	\\
Nov. 20.699	& 1.983 &7.271 &0.708&0.495 &1.021 &	\\
Nov. 21.706	&2.990 &7.554 &0.620&0.572 &1.144 &	\\
Nov. 23.687	& 4.971 &7.455 &0.524&0.567 &1.144 &	\\
Nov. 24.687	& 5.971 &7.401 &0.545&0.542 &1.107 &	\\
Nov. 25.690	& 6.974 &7.473 &0.578&0.550 &1.101 &	\\
Nov. 28.686 & 9.970 &7.327 &0.581&0.549 &1.053 &	\\
Dec. 2.681	& 13.965 &7.508 &0.583&0.550 &1.109 &	\\
Dec. 3.682	& 14.966 &7.907 &0.484&0.668 &1.212 &	\\
Dec. 14.688	& 25.972 &8.250 &0.450&0.759 &1.251 &	\\
Dec. 18.692	& 29.976 &8.190 & & &1.097 &	\\
2010        &       &      &     &      &      & \\
Feb. 2.221	& 75.505 &9.607 &0.351& &1.402 &	\\
Feb. 8.212	& 81.496 &9.861 &0.301&1.053 &1.436 &	\\
Feb. 15.204	& 88.488 &10.373 &0.141&1.154 &1.561 &	\\
Feb. 27.196	& 100.480 &11.178&0.375&1.477 &1.764&	\\
Mar. 6.163	& 107.447 &11.684&0.501&1.796 &2.003 &	\\
Mar. 14.172	& 115.456 &11.143&0.152&1.412 &1.588 &	\\
Mar. 16.165	& 117.449 &11.229&0.175&1.500 &1.591 &	\\
Mar. 24.168	& 125.452 &11.162&0.320&1.461 &1.465 &	\\
Mar. 27.172	&128.456 &11.314& & &1.375 &	\\
Apr. 2.141	& 134.425 &11.306&0.297&1.412 &1.259 &	\\
Apr. 7.119	& 139.403 &11.327&0.288&1.409 &1.244 &	\\
Apr. 15.110	& 147.394 &11.405 &0.290&1.417 &1.182 &	\\
Apr. 20.128	& 152.412 &11.466&0.290&1.375&1.127&	\\
Apr. 28.094	& 160.378 &11.439&0.292&1.313 &1.037 &	\\
May. 17.020	& 179.305 &11.411&0.315&1.117 &0.823 &	\\
May. 18.089	& 180.374 &11.377&0.387&1.074 &0.794 &	\\
May. 19.067	& 181.352 &11.401&0.352&1.087 &0.770 &	\\
May. 21.040	& 183.324 &11.371&0.376&1.005 &0.742 &	\\
May. 25.004	& 187.289 &11.424&0.310&1.090 &0.740 &	\\
May. 28.008	& 190.293 &11.395&0.386&0.994 &0.678 &	\\
Jun. 4.004	& 197.289 &11.413&0.359&0.920 &0.653 &	\\
Jun. 10.978	& 204.262 &11.447 &0.381&0.877 &0.557 &	\\
Jun. 21.950	& 215.235 &11.448&0.516&0.711& 0.416&	\\
Jun. 28.918	& 222.202 &11.607&0.293&0.816 & 0.464&	\\
Jul. 8.947	& 232.232 &11.584 &0.501&0.641 &0.315 &	\\
Jul. 16.906	& 240.191 &11.680&0.455&0.657& 0.268 &	\\
Jul. 24.914	& 248.199 &11.677&0.602&0.514 &0.170 &	\\
Aug. 1.962	& 256.246 &11.770&0.559&0.513 &0.171  &	\\
Aug. 16.905	& 271.189 &11.841&0.666&0.396 &0.052&	\\
Aug. 22.883	& 277.167 &11.925&0.627&0.404 & 0.012&	\\
Aug. 31.866	& 286.150 &11.937&0.759&0.302 & -0.044&	\\
Sep. 10.859	& 296.080 &11.999&0.796&0.238 & -0.111&	\\
Sep. 20.774	& 305.995 &12.085&0.772&0.212 & -0.136&	\\
Sep. 29.844	& 315.065 &12.141&0.774&0.193 & -0.139&	\\
Oct. 10.752 & 325.973 &12.188&0.843&0.139 & -0.211&	\\
Oct. 19.777	& 334.998 &12.225&0.878&0.098 & -0.230&	\\
Oct. 27.762	& 342.984 &12.267&0.915&0.097 & -0.220&	\\
Nov. 12.744	& 358.965 &12.387& & & &	\\
\end{tabular}
\end{table}
\begin{table}
\begin{tabular}{lccccccc}
2011        &       &      &     &      &      & \\
Feb. 10.212 & 448.433 &12.773&   &       & -0.557&	\\
Feb. 23.190	& 461.411 &12.813&1.040&-0.234 &-0.642&	\\
Mar. 20.151	& 486.372 &12.933&1.066&-0.287 &-0.624&	\\
Apr. 13.141	& 511.362 &13.029&  &   & -0.858&	\\
Apr. 18.087	& 516.308 &13.128&0.901&-0.187 &-0.683&	\\
Apr. 21.088& 519.309 &13.109&0.957&-0.256 &-0.672&	\\
\hline
\end{tabular}
\end{table}


\begin{table}
\caption{Log of the Varese optical spectroscopy of V496 Sct. The
date of optical maximum is taken as 2009 November 18.716 UT.}
\centering

         \begin{tabular}{@{~}rcc@{~~}c@{~~}cc@{~}}
            \hline
\multicolumn{1}{c}{date} &  UT   & expt  &    disp  & $\lambda$ range & Days since  \\
                         &       & (sec) & (\AA/pix)& (\AA)           &    optical maximum   \\
            \hline
2009 11 09 & 18:26  & 1500 & 2.12           & 3960-8600       & -8.955 \\
        10 & 18:15  & 3600 & echelle        & 3950-8640       & -7.960 \\
        12 & 17:06  &  900 & 2.12           & 3955-8595       & -6.005 \\
        12 & 17:50  & 1800 & echelle        & 3880-8640       & -5.987\\
        19 & 16:52  &  360 & 2.12           & 3955-8590       & 0.972 \\
        19 & 17:57  & 3600 & echelle        & 3950-8650       & 1.016 \\
        21 & 16:53  &  420 & 2.12           & 3965-8610       & 2.973\\
        21 & 17:54  & 3600 & echelle        & 4180-8650       & 3.015\\
        24 & 17:46  & 2700 & echelle        & 3955-8645       & 6.012\\
        28 & 17:07  & 2700 & echelle        & 4020-8655       & 9.995\\
     12 01 & 17:23  &  900 & echelle        & 3955-8645       & 13.002 \\
        01 & 17:58  &  900 & 2.12           & 3945-8590       & 13.017 \\
        05 & 17:09  & 3600 & echelle        & 3950-8645       & 16.996\\
        10 & 17:04  & 1800 & echelle        & 3950-8640       & 21.994 \\
        10 & 17:34  &  300 & 2.12           & 3945-8597       & 22.007 \\
2010 03 13 & 04:43  & 1020 & 4.24           & 3800-8385       & 114.469 \\
     04 28 & 02:39  & 2400 & 2.12           & 3965-8600       & 160.384\\
     06 06 & 01:13  & 3600 & 2.12           & 3975-8615       & 199.331 \\
     06 22 & 23:07  & 3600 & 2.12           & 3975-8600       & 216.245 \\
     07 15 & 23:24  & 3600 & echelle        & 3950-8640       & 239.252 \\
     10 05 & 19:16  & 3600 & 2.12           & 3925-8565       & 321.082 \\
2011 04 19 & 02:47  & 5400 & 2.12           & 4000-8640       & 516.387 \\
            \hline
\end{tabular}
\label{tb:tab2}
\end{table}

\section{Results}
\subsection{The optical light curve}
\subsubsection{The pre-maximum rise, outburst luminosity, reddening and distance}
The optical light curve based on Table 2 is presented in Fig. 1. There is a good photometric coverage of the nova's rise to maximum which lasts for almost 10 days culminating in a peak brightness of
$V_{max}$ = 7.07 on 2009 November 18.716 UT. The early decline after the maximum was observed till mid -December and subsequently the solar conjunction of V496 Sct resulted in lack of its observational coverage till early 2010 February. We determine the rate of decline by doing a least square regression fit to the post maximum light curve and estimate $t_2$ to be 59 $\pm$ 5 d. The estimated value of $t_2$ makes V496 Sct as one of the moderately fast Fe\,{\sc ii} class of novae in recent years.
 V496 Sct is one of the large amplitude novae observed in recent years with $\bigtriangleup$R $\ge$ 13.5 magnitudes (Guido \& Sostero 2009). These observed values of the amplitude and $t_2$  for V496 Sct put it close to the upper limit in the observed spread of the amplitude versus decline rate plot for classical novae presented by Warner (2008, Fig. 2.3).
Using the maximum magnitude versus rate of decline (MMRD) relation of della Valle \& Livio (1995), we determine the absolute magnitude of the nova to be $M_V = -7.0$ $\pm$ 0.2. The reddening is derived using the intrinsic colors of novae at peak brightness, namely $(B-V)$ = 0.23 $\pm$ 0.06, as derived by van den Bergh $\&$ Younger (1987). We have used our optical photometry data to calculate $E(B-V)$. The observed $(B-V)$ = 0.797 $\pm$ 0.014 results in $E(B-V)$ = 0.57 $\pm$ 0.06. We have also estimated $E(B-V)$ using the interstellar lines and diffuse interstellar band (DIB) registered in our high resolution optical spectra. The Na I line is composed of at least five independent components, of which three are well isolated and the remaining two are blended. Following Munari \& Zwitter (1997) we estimate a total value of $E(B-V)$ = 0.65 from these five components. The interstellar line of K I, though underexposed in the observed spectra, gives a value of $E(B-V)$ $\sim$ 0.60 and the DIB $\lambda$6614 gives $E(B-V)$ = 0.65. These estimates of interstellar reddening are in good agreement with each other, and in the rest of this paper we adopt $E(B-V)$ = 0.57 $\pm$ 0.06 and $A_V$ = 1.77 $\pm$ 0.06 for a standard reddening law $R$ = 3.1. In their study of the spatial distribution of the interstellar extinction, Neckel $\&$ Klare (1980) have shown that close to the direction of V496 Sct, $A_V$ shows a value of $\sim$ 1.8 mag around 3 kpc and moderate value of $A_V$ estimated by us appears reasonable. By using (MMRD) relation of della Valle \& Livio (1995) and taking the value of $E(B-V)$ = 0.57 $\pm$ 0.06 we obtain the distance $d = 2.9$ $\pm$ 0.3 kpc to the nova. By using the relations for the blackbody angular diameter and temperature, expansion rate for the ejecta and distance to the nova given by Ney \& Hatfield (1978) and Gehrz (2008) respectively, we estimate a value $\sim$ 9 kpc for the distance to the nova. This value is more than 3 times the value estimated by (MMRD) relation of della Valle \& Livio (1995). A likely reason for this discrepancy is the behaviour of the pseudo-photosphere as a grey-body with reduced emissivity in the fireball phase as seen earlier in the case of V1280 Sco (Das et al. 2008) and V5579 Sgr (Raj, Ashok $\&$ Banerjee 2011).

\subsubsection{The nature of the light curve}

A classification system for the optical light curves for novae, based on a large sample of American Association of Variable Star Observers (AAVSO), has been presented by Strope et al. (2010). Their classification system defines seven classes based on the time to decline by 3 mag from the peak, t$_{3}$, and the shape of the light curve. The shape of the optical light curves of V496 Sct presented in Fig. 1 has all the characteristics of D class of nova. The early decline following the rise to the maximum is interrupted by fast decline around 90 days after the outburst reaching minimum brightness close to 120 days near the center of the dust dip. The brightness recovered to a value below the original decline. Thus the classification of the optical light curve for V496 Sct is D(90), as the estimated value of t$_{3}$ is $\sim$ 90 days for V496 Sct.

 \begin{figure*}
 \includegraphics [width=5.5in,height=6.0in,clip]{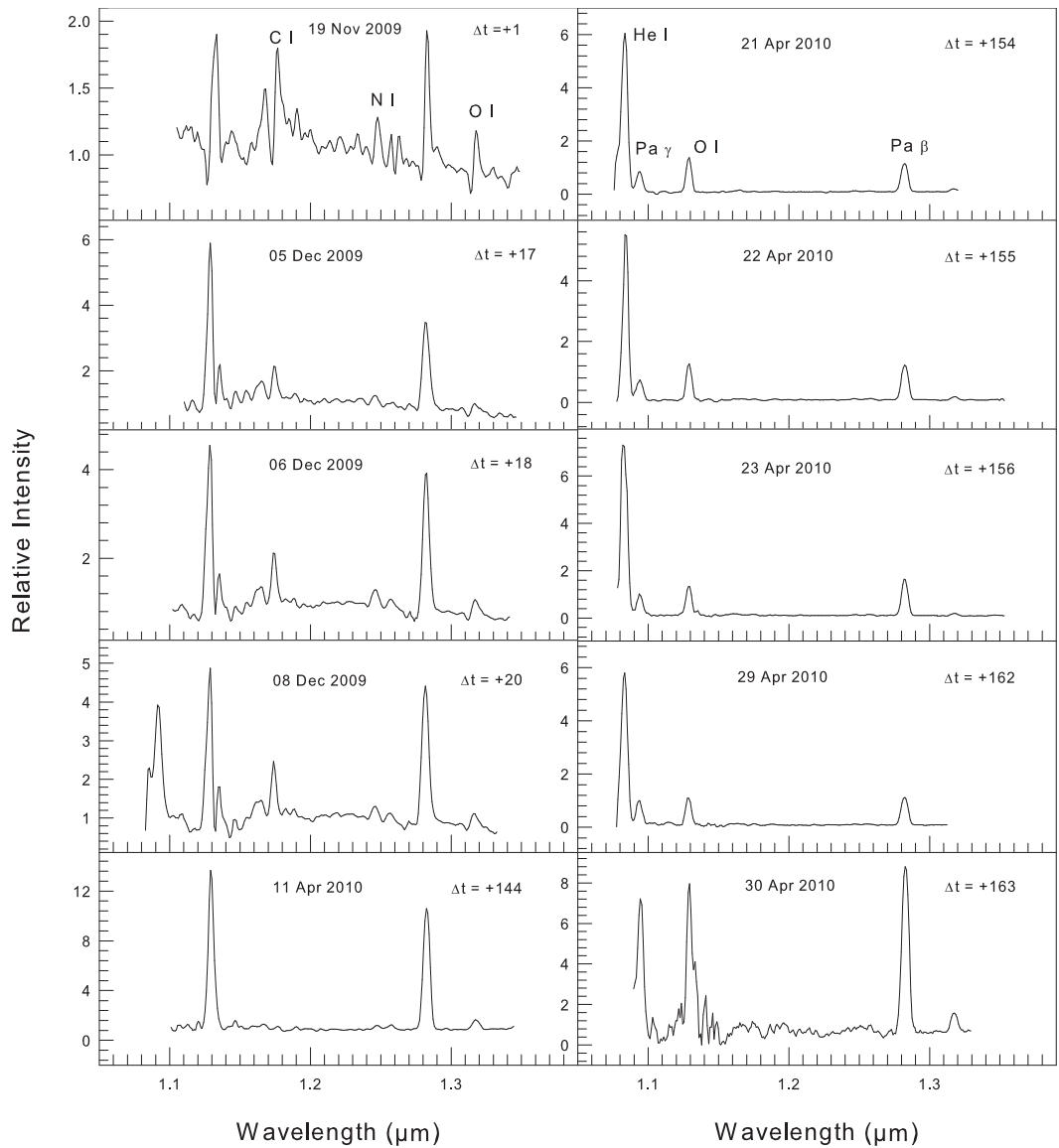}
  \caption[]{ The $J$ band spectra of V496 Sct are shown at different epochs. The relative intensity is normalized to unity at 1.25 $\mu m$. The time from optical maximum are given for each spectrum.}
 \label{fig2}
 \end{figure*}


\subsection{Line identification, evolution and general characteristics of the $JHK$ spectra}
 The $JHK$ spectra are presented in Figs. 2 to 4 respectively; the observed line list is given in
 Table 4. The infrared observations presented here cover all the phases with the first J band spectrum taken on 2009 November 19 very close to the visual maximum. This J band spectrum is dominated by the lines of
 H\,{\sc i}, N\,{\sc i}, C\,{\sc i} and O\,{\sc i} all displaying deep P Cygni profiles. The full width at half maximum (FWHM) of the emission and the absorption components of P$\beta$ line are 700 km s$^{-1}$ and 270 km s$^{-1}$ respectively. The absorption component is blue shifted by 960 km s$^{-1}$ from the emission component. The next set of spectra taken beginning from 2009 December 3 show the disappearance of P Cygni profiles and predominant emission components for all the lines. The typical FWHM of the H\,{\sc i} lines are
1230 $\pm$ 50 km s$^{-1}$. The ratio of the observed strength of the O\,{\sc i} lines, W (1.1287)/W (1.3164) $\sim$ 3 indicates that  Ly $\beta$ fluorescence is the dominating pumping mechanism and this is corroborated by the strong O\,{\sc i} 8446 line seen in the optical spectra discussed later. A noticeable feature of these early spectra is the  presence of lines due to Na\,{\sc i} and Mg\,{\sc i}. In the spectra taken on 2009 December 5 the Na\,{\sc i} lines at 1.1404$\mu$m, 2.1452$\mu$m,
2.2056$\mu$m and 2.2084$\mu$m and  Mg\,{\sc i} lines at 1.1828$\mu$m, 1.5040$\mu$m, 1.5749$\mu$m and 1.7109 ${\mu}$m are clearly seen. In an earlier study of V1280 Sco, Das et al. (2008) had suggested
that the presence of spectral lines of low ionization species like Na\,{\sc i} and Mg\,{\sc i} in the early spectra are indicators of low temperature zones conducive to dust formation in the nova ejecta and this is very well borne out in the case of V496 Sct. We would like to point out the presence of a large number of strong lines of neutral carbon seen in the $JHK$ bands. These are typical of Fe\,{\sc ii} type nova as seen in the case of V2615 Oph (Das, Banerjee $\&$ Ashok 2009) and V5579 Sgr (Raj, Ashok $\&$ Banerjee 2011).

The most interesting spectral features seen in the spectra of V496 Sct taken in 2009 December are the prominent first overtone CO bands in the K band and they are discussed in the following subsection 3.3.
The last spectra, before V496 Sct became inaccessible due to its conjunction with the sun, was taken on 2009 December 9.

The set of spectra taken in 2010 April, after V496 Sct emerged from its solar conjunction, show strong He I lines at 1.0830${\mu}$m in the J band and 2.0581${\mu}$m in the K band. The He I 1.0830${\mu}$m line exceeds in strength compared to H I lines indicating higher excitation conditions in the nova ejecta. The other weaker He I lines at 1.7002${\mu}$m and 2.1120-2.1132${\mu}$m are also seen. The rising continuum seen in the spectra taken on 2010 April 11 indicates that the dust detected by Russell et al. (2010) on 2010 February 10 is still present.

 \begin{figure*}
 \includegraphics[width=5.5in,height=6.0in, clip]{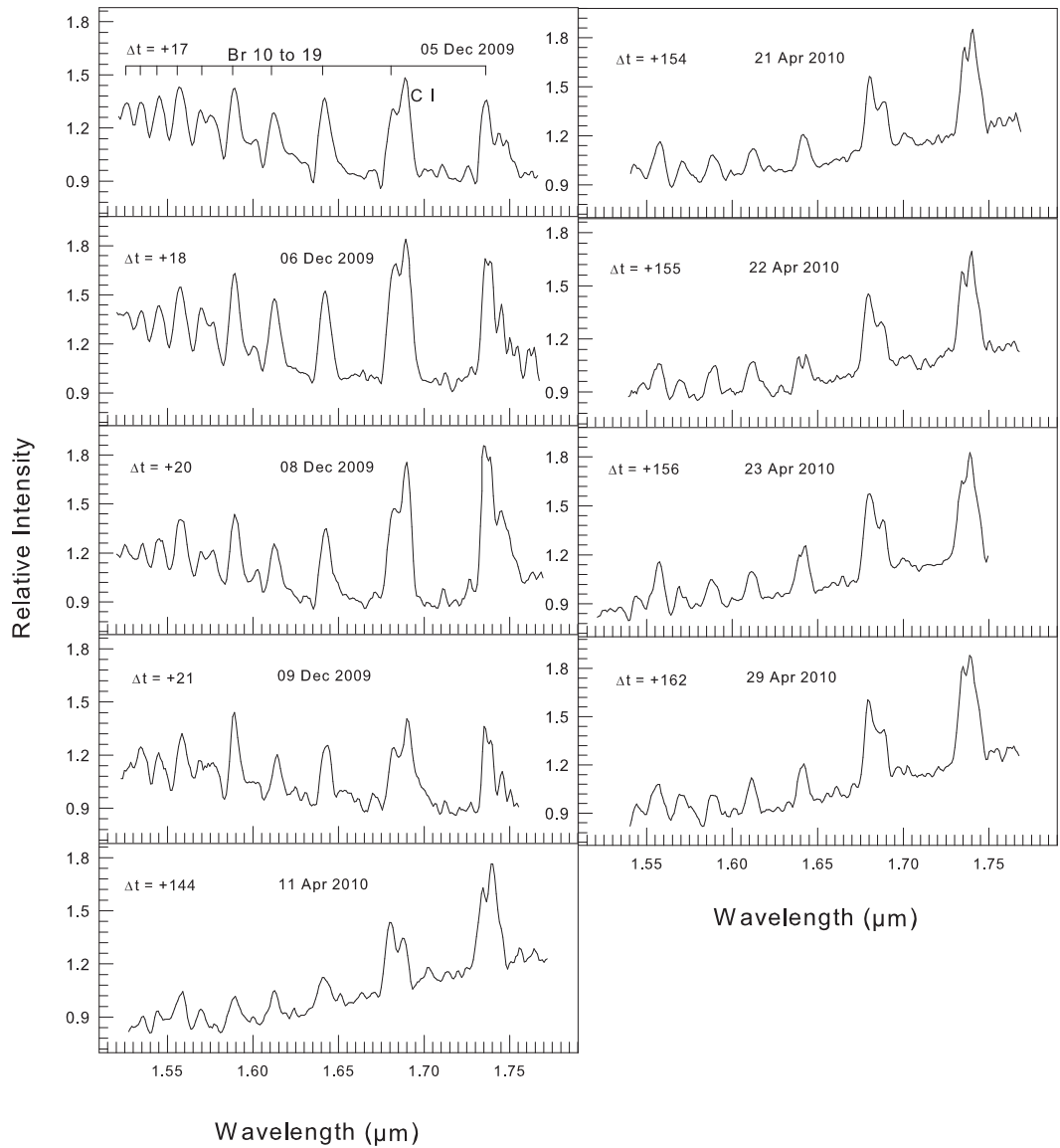}
  \caption[]{The $H$  band spectra of V496 Sct are shown at different epochs. The
   relative intensity is normalized to unity at 1.65 $\mu m$. The time from optical maximum are given for each spectrum.}
  \label{fig3}
  \end{figure*}
%
%
%
  \begin{figure*}
  \includegraphics[width=5.5in,height=7.2in, clip]{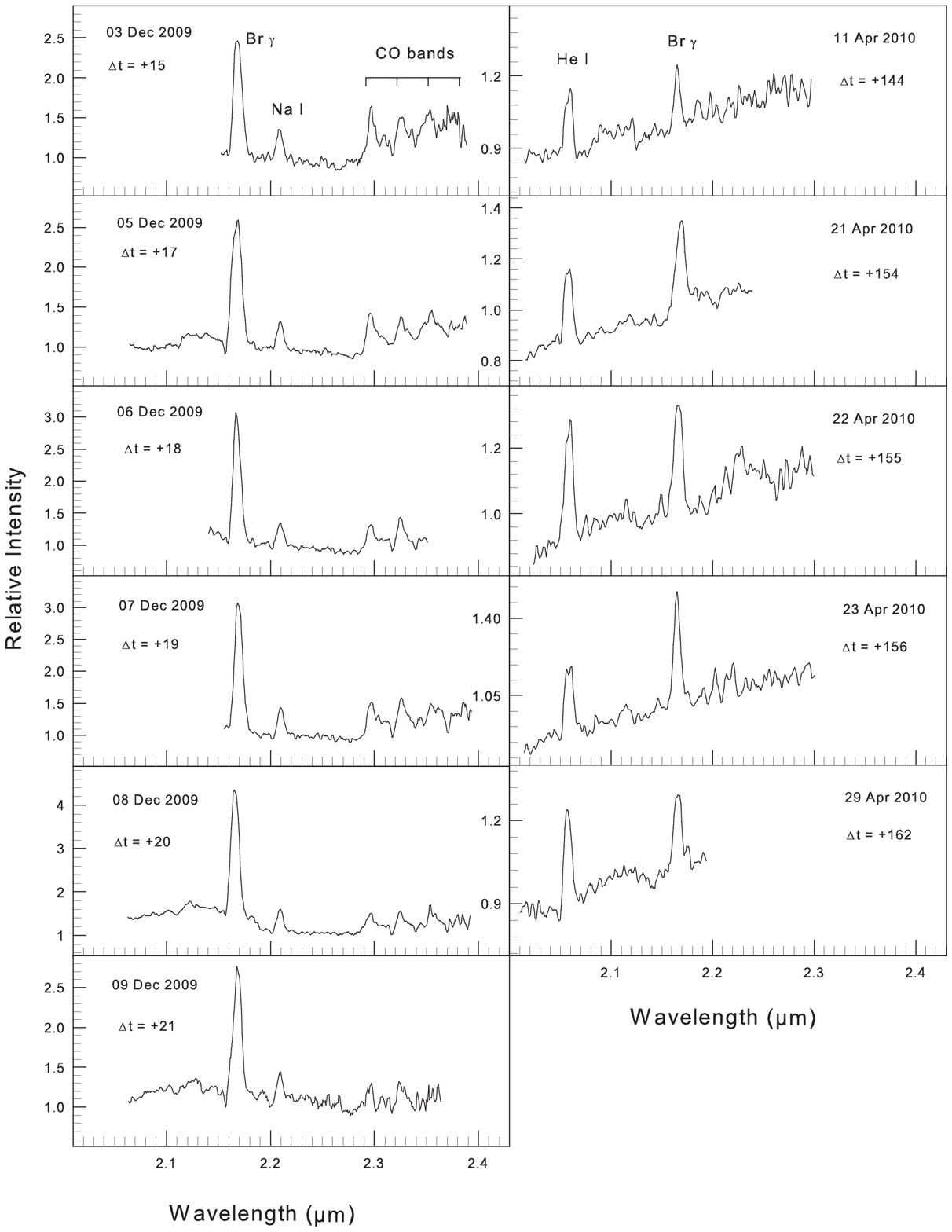}
 \caption[]{The $K$  band spectra of V496 Sct are shown at different epochs. The
   relative intensity is normalized to unity at 2.2 $\mu m$. The time from optical maximum are given for each spectrum.}
     \label{fig4}
  \end{figure*}
\begin{table}
\caption[]{A list of the lines identified from the $JHK$ spectra. The
  additional lines contributing to the identified lines are listed.}
\begin{tabular}{llll}
\hline\\
Wavelength & Species  & Other contributing  \\
(${\mu}$m) & &lines and remarks   \\
\hline \\

1.0830   & He\,{\sc i}         &      \\
1.0938   & Pa $\gamma$        &      \\
1.1287          & O\,{\sc i} &               \\
1.1404   & Na\,{\sc i}        &    C\,{\sc i} 1.1415  \\
1.1600-1.1674   & C\,{\sc i}  & strongest lines at 1.1653,\\
                &             &           1.1659,1.16696    \\
1.6872          & Fe\,{\sc ii}&                             \\
1.1746-1.1800   & C\,{\sc i}  & strongest lines at 1.1748,  \\
                &             &          1.1753,1.1755      \\
1.1828          & Mg\,{\sc i} &               \\
1.1880          & C\,{\sc i} &               \\
1.2074          & N\,{\sc i} &               \\
1.2187,1.2204          & N\,{\sc i} &               \\
1.2249,1.2264          & C\,{\sc i} &               \\
1.2329          & N\,{\sc i} &               \\
1.2382          & N\,{\sc i} &               \\
1.2461,1.2469 & N\,{\sc i} & blended with O\,{\sc i} 1.2464 \\
1.2562,1.2569 & C\,{\sc i} & blended with O\,{\sc i} 1. 2570 \\
1.2818   & Pa $\beta$         &     \\
1.2950   & C\,{\sc i}         &     \\
1.3164   & O\,{\sc i}         &     \\
1.5040   & Mg\,{\sc i}        &  blended with Mg\,{\sc i} 1.5025,\\
         &                    &         1.5048 \\
1.5256   & Br 19              &           \\
1.5341   & Br 18              &           \\
1.5439   & Br 17              &           \\
1.5557   & Br 16              &           \\
1.5701   & Br 15              &           \\
1.5749   & Mg\,{\sc i}        & blended with Mg\,{\sc i} 1.5741,  \\
         &                    &           1.5766,C\,{\sc i} 1.5788 \\
1.5881   & Br 14              &  blended with C\,{\sc i} 1.5853    \\
1.6005   & C\,{\sc i}         &           \\
1.6109   & Br 13              &           \\
1.6407   & Br 12              &           \\
1.6806   & Br 11              &           \\
1.6890   & C\,{\sc i}         &           \\
1.7002   & He\,{\sc i}         &           \\
1.7045   & C\,{\sc i}         &           \\
1.7109   & Mg\,{\sc i}        &               \\
1.7234-1.7275 & C\,{\sc i}    & several C\,{\sc i} lines  \\
1.7362   & Br 10              &  affected by C\,{\sc i} 1.7339 line    \\
1.7449 & C\,{\sc i}           &           \\
1.7605-1.7638 & C\,{\sc i}    &           \\
2.0581 & He\,{\sc i}          &           \\
2.1120,2.1132   & He\,{\sc i}         &           \\
2.1156-2.1295 & C\,{\sc i}    &           \\
2.1452 & Na\,{\sc i}          &           \\
2.1655   & Br $\gamma$        &           \\
2.2056,2.2084 & Na\,{\sc i}   &           \\
2.2156-2.2167 & C\,{\sc i}    &           \\
2.29-2.40           & CO               & $\Delta$v=2 bands             \\
2.2906   		    & C \,{\sc i}      	& 				\\
2.3130              & C \,{\sc i}       &               \\
\hline
\end{tabular}
\label{table4}
\end{table}

\subsection{Modeling and evolution of the CO emission}

We adopt the model developed in our earlier work on V2615 Oph (Das, Banerjee \& Ashok 2009) to characterize the CO emission. Briefly, in this model the CO gas is considered to be in thermal equilibrium with the same temperature for calculating the level populations of rotation and vibration bands. It is assumed that the rotational levels are gaussian in shape. In addition to $^{12}$C $^{16}$O the other isotopic species included in the calculations is $^{13}$C $^{16}$O. The isotopic species like $^{12}$C $^{17}$O and $^{12}$C $^{18}$O are not considered as they are expected to have low abundances. The model luminosity E which is in units of erg s$^{-1}$ is converted to erg cm$^{-2}$ s$^{-1}$$\mu$m$^{-1}$ by dividing with 4$\pi$d$^{2}$ where d is the distance to the source and scaling to a unit wavelength. The peak intensities of the vibration bands are analytically determined such that the integrated area under the curve matches the expected observed quantity E/4$\pi$d$^{2}$. An appropriate continuum determined from the K band photometry for a particular date is added to the model CO emission so that it can be compared with the observed CO emission bands. The input parameters to the model are the total mass of the CO gas (M$_{CO}$), the $^{12}$C/$^{13}$C ratio denoted as a constant $\alpha$ and the gas temperature T$_{CO}$. For a given set of values for M$_{CO}$,  $\alpha$, T$_{CO}$ and d the CO flux estimated from the model is an absolute quantity. The representative model spectra matching the observed data for 2009 December 5 and 7 are shown in Fig. 5.

The best fit model spectra to the observed data are obtained by varying the input parameters M$_{CO}$,  $\alpha$, T$_{CO}$. The expected changes to the model spectra by varying these input parameters may qualitatively be summarized as follows. The increase in M$_{CO}$ enhances the absolute level of the CO emission while the increase in T$_{CO}$ changes the relative intensities of different vibrational bands in addition to changing the absolute level of the emission. The CO emission is assumed to be optically thin. The C I lines at 2.2906 and 2.3130 $\mu$m and Na I lines at 2.3348 and 2.3379 $\mu$m are also likely to be present in the spectral region covered by the CO emission giving rise to some deviations between the best model fit and the observed spectra. In addition, since a comparison of the relative strengths of the vibrational bands within the first overtone allows the gas temperature to be determined, we are handicapped by being able to detect only three of the bands ($\nu$ = 2-0, 3-1, 4-2). Within these constraints our formal model fits for 2009 December 5 and 7, yield temperatures of 4000 $\pm$ 500 K and 3600 $\pm$ 500 K respectively with a reasonably similar range in mass of M$_{CO}$ = 1.5$-$2 $\times$ 10$^{-8}$ M$_\odot$. The representative fits for 2009 December 5 and 7 are shown in Fig. 5. The model calculations also show that the $v = 2-0$ bandhead of $^{13}$CO at 2.3130 $\mu$m becomes discernibly prominent if the $^{12}$C/$^{13}$C  ratio is $\leq$ 1.5. As this spectral feature is not clearly detected in our observed spectra, we place a lower limit of $\sim$ 1.5 for the $^{12}$C/$^{13}$C ratio. However we add a cautionary note that the signal to noise ratio in the region of $^{12}$CO and $^{13}$CO bands is about 15-20 and better quality spectra may permit a more accurate determination of the CO gas parameters. The lower limit of $\sim$ 1.5 for the $^{12}$C/$^{13}$C ratio reported here is the lowest value till date among the novae that have displayed the first overtone bands of CO.

It may be helpful to compare the observed $^{12}$C/$^{13}$C ratio in V496 Sct with the values for other novae that have displayed the first overtone bands of CO and also the model predicted values. The observed values for the $^{12}$C/$^{13}$C ratio available in the literature are $^{12}$C/$^{13}$C $\geq$ 5 in V705 Cas (Evans et al. 1996), $^{12}$C/$^{13}$C $\geq$ 3 in NQ Vul (Ferland et al. 1979), $^{12}$C/$^{13}$C $\geq$ 2 in V2615 Oph (Das, Banerjee \& Ashok, 2009), $^{12}$C/$^{13}$C $\simeq$ 2.9 in V842 Cen (Wichmann et al. 1991) and $^{12}$C/$^{13}$C $\simeq$ 1.2 in V2274 Cyg (Rudy et al. 2003). The thermonuclear runaway (TNR) responsible for the nova outburst is one of the important source for the production of $^{13}$C isotopes (Starrfield et al. 1972; Starrfield, Sparks \& Truran 1974 and Romano \& Matteucci 2003). Hajduk et al. (2005) have pointed out that the outburst of born-again giants like V4334 Sgr are another source for $^{13}$C isotopes in the galaxy; a low value for the ratio $^{12}$C/$^{13}$C = 5 was observed by Pavlenko et al. (2004) in V4334 Sgr. In the hydrodynamical models of nova outbursts the $^{12}$C/$^{13}$C ratio will depend on parameters like the mass of the underlying white dwarf, the accretion history and mixing of the accreted material from the companion star with the surface material of the white dwarf (Jose \& Hernanz 1998; Starrfield et al. 1997; Yaron et al. 2005). The estimated lower limits as well as the observed values for $^{12}$C/$^{13}$C ratio in case of novae discussed about indicate that $^{13}$C is possibly not synthesized at the high values predicted by these theoretical models.

  \begin{figure}
\includegraphics[width=3.0in,height=3.0in, clip]{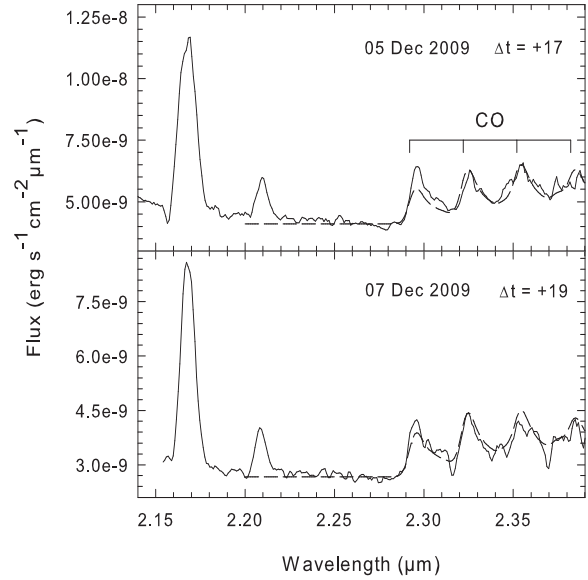}
  \caption[]{ The model fits are shown as dashed lines to the observed first overtone CO bands in V496 Sct for 2009 December 5 and 7. The fits are made for a constant CO mass of 2e-8 M$_\odot$ on both the days while the temperature of the gas T$_{CO}$ is 4000 K and 3600 K respectively. The time from optical maximum are given for each spectrum. }
  \label{fig5}
  \end{figure}
In the novae mentioned above that displayed CO bands, the estimated M$_{CO}$ ranges from 2.8 $\times$ 10$^{-10}$ M$_\odot$ (V705 Cas) to 3 $\times$ 10$^{-8}$ M$_\odot$ (V2615 Oph). The present value of M$_{CO}$ = 1.5$-$2 $\times$ 10$^{-8}$ M$_\odot$ for V496 Sct lies within this range and is similar to the CO mass determined in V2615 Oph.

\begin{table}
\caption[]{A list of the prominent emission lines identified from the optical spectra.}
\begin{tabular}{llll}
\hline\\
Wavelength & Species &   \\
(A) & &                 \\
\hline \\

3970   & Ca\,{\sc ii} and H$\epsilon$      &    \\
4101   & H$\delta$        &  \\
4129   &   Fe\,{\sc ii}(27)        &     \\
4173   & Fe\,{\sc ii}(27) & \\
4233   & Fe\,{\sc ii}(27)  &  \\
4303          &  Fe\,{\sc ii}(27)  &        \\
4340   & H$\gamma$  &\\
4351 & Fe\,{\sc ii}(27) &\\
4363 & [O\,{\sc iii}] & \\
4555 & Fe\,{\sc ii}(37) & \\
4586   & Fe\,{\sc ii}(38) &  \\
4634   & N\,{\sc iii}        & \\
4649   & O\,{\sc ii}         &  \\
4686   & He\,{\sc ii}        &  \\
4861   & H$\beta$              &       \\
4924   & Fe\,{\sc ii}(42)             &          \\
4959   & [O\,{\sc iii}]              &           \\
5007   & [O\,{\sc iii}]             &         \\
5018   & Fe\,{\sc ii}(42)              &        \\
5046   & Si\,{\sc ii}   & \\
5159   & [Fe\,{\sc vi}] + [Fe\,{\sc vii}]       &\\
5169   & Fe\,{\sc ii} + Mg\,{\sc i}          &       \\
5235   &      Fe\,{\sc ii}(49)           &           \\
5270   & [Fe\,{\sc iii}]  &  \\
5276   & Fe\,{\sc ii}(49+48)        &           \\
5309   & [Ca\,{\sc v}]             &   \\
5316   & Fe\,{\sc ii}(49)         &       \\
5361   & Fe\,{\sc ii}(48)         &     \\
5415   & He\,{\sc ii}    & \\
5535   & Fe\,{\sc ii}(55) + N\,{\sc ii}        &   \\
5676   & N\,{\sc ii}  & \\
5755 & [N\,{\sc ii}](3)    & \\
5876   & He\,{\sc i}              &    \\
5890 & Na\,{\sc i}           &       \\
5909 & Fe\,{\sc ii}    &          \\
5942 & N\,{\sc ii}(28)          &        \\
5991 & Fe\,{\sc ii}(46)    &       \\
6086 & [Ca\,{\sc v}] + [Fe\,{\sc vii}]   &   \\
6084 & Fe\,{\sc ii}(46)          &     \\
6157   & O\,{\sc i}       &         \\
6243 & Fe\,{\sc ii} + N\,{\sc ii}   &          \\
6300 & [O\,{\sc i}]    &         \\
6347           &          Si\,{\sc ii}(2)       &         \\
6363   		    & [O\,{\sc i}]      	&			\\
6419              & Fe\,{\sc ii}(74)         &            \\
6431              &       Fe\,{\sc ii}(40)  &            \\
6456              & Fe\,{\sc ii}         &            \\
6563              & H$\alpha$       &            \\
6678              & He\,{\sc i}      &          \\
6726              &        O\,{\sc i}(2)  &             \\
7065              & He\,{\sc i} &  \\
7139       &    [Ar\,{\sc iii}]   & \\
7234              &  C\,{\sc ii}(3, blend of 7231 and 7236)    &            \\
7330              & [O\,{\sc ii}]       &           \\
7774              & O\,{\sc i}       &         \\
8446              & O\,{\sc i}       &       \\
8498              & Ca\,{\sc ii} triplet   &                 \\
8542              &  Ca\,{\sc ii} triplet      &             \\

\hline
\end{tabular}
\label{table4}
\end{table}

\subsection{Line identification, evolution and general characteristics of the optical spectra}
The optical spectra presented here cover the pre- maximum rise, the optical
maximum brightness, the early decline and the nebular phase.  There are very few novae for which the spectral evolution before the maximum brightness has been documented: V1280 Sco (Naito et al.  2012, Kuncarayakti et al. 2008) and V2615 Oph (Munari et al. 2008).
The spectral evolution of V496 Sct during the pre-maximum, maximum and
optically thick branch of the decline phase is presented in Fig. 6, while
the subsequent evolution during the optically thin and nebular phase is
covered by Fig. 7. Fig. 8 documents the complex evolution of profiles
of Fe II $\lambda$5018 line, and Fig. 9 the temporal evolution of the velocity of the
absorption components of Fe II $\lambda$5018 line. A summary of emission lines identified in the optical spectra is provided in Table~5, and de-reddened fluxes (according to $E_{B-V}$ = 0.57 and a $R_V$=3.1 reddening law) of the prominent emission lines relative to H$\beta$ are given in Table 6 for some representative dates.

\subsubsection {Pre-maximum rise and optical maximum}
There is considerable interest in studying the spectral evolution during the pre-maximum rising phase of classical novae, an evolutionary phase rarely observed. The pre-maximum spectra of V496 Sct in Fig. 6 are characterized by emission lines confined to just FeII (multiplets 27, 28, 37, 38, 42, 48, 49, 55 and 74) and hydrogen Balmer series, with feeble OI 7772, 8446.

During the rise toward maximum the baricentric and terminal velocity of
P Cygni absorptions, and the FWHM of both absorption and emission components
declined with time (see Fig. 6). The terminal and core velocity of P Cygni absorption component of H$\alpha$ line in Fig 6. declined from $-$2000 and $-$700 km $s^{-1}$ on day $-$9 to $-$1200 and $-$600 km $s^{-1}$ on
day $-$6, while at the same time the FWHM of the absorption and emission
components declined from 1000 to 700 km $s^{-1}$. The P Cygni components essentially vanished at the time of maximum brightness. When they reappeared later into early decline they were much sharper (FWHM=250 km $s^{-1}$) and blue-shifted (core velocity = $-$1350 km $s^{-1}$) than at pre-maximum.

The high resolution Echelle spectra listed in Table 3 show that the broad and single P Cygni
absorption components observed on 2009 November 10 and 12 of H I and Fe II
multiplets are replaced by two narrow components on 2009 November 19. The two absorption
components in case of Fe II 5018, shown in Fig. 8, are located at
heliocentric velocities of -785 and -360 km $s^{-1}$.

 \begin{figure*}
\includegraphics[width=4.0in,height=6.0in,angle=270,clip]{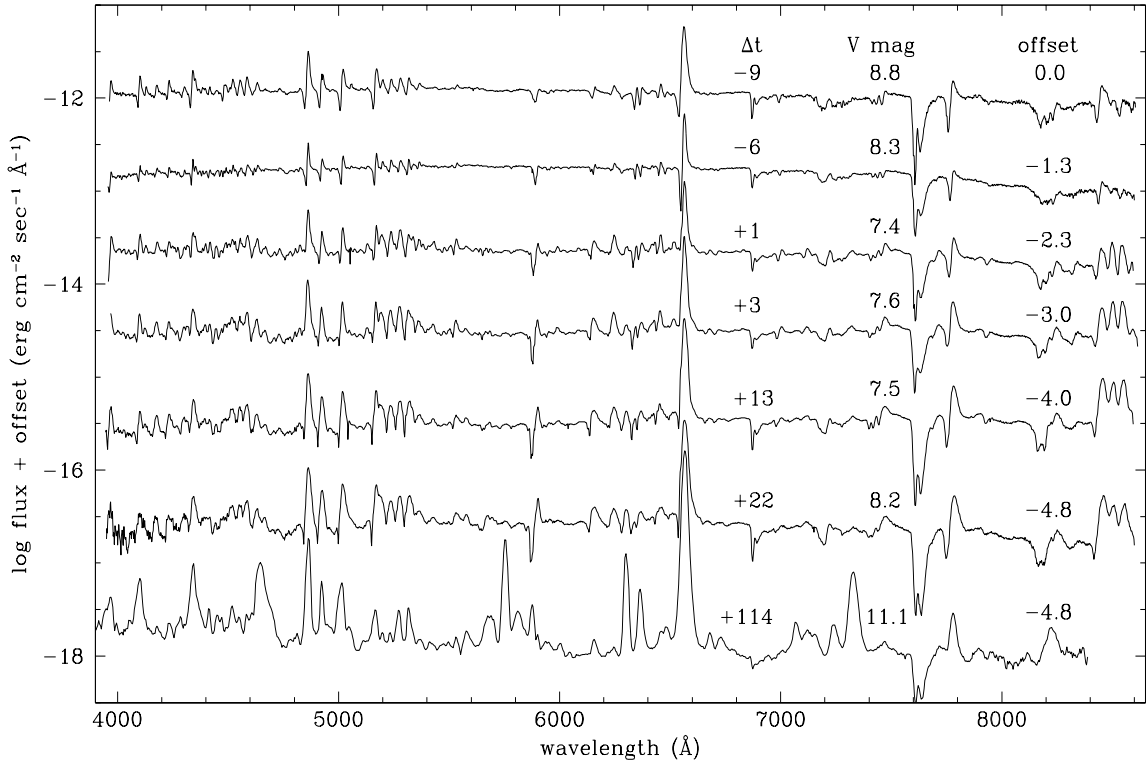}
\caption[]{ Low-res spectroscopic evolution of V496 Sct from pre-maximum to the end
 of the optically thick branch (pre-nebular stage) of the light-curve. For the
  first month of the evolution the emission lines are mostly due to Fe II and H I,
  with also O I 7772, 8446, Na I 5893, and Ca II 8498, 8542. The last spectrum in
  plotted for commonality also in Fig. 7. The time from optical maximum,
  the $V$ band magnitude and the offset in log flux are given for each spectrum.}
  \label{fig6}
  \end{figure*}

 \begin{figure*}
\includegraphics[width=4.0in,height=6.0in,angle=270,clip]{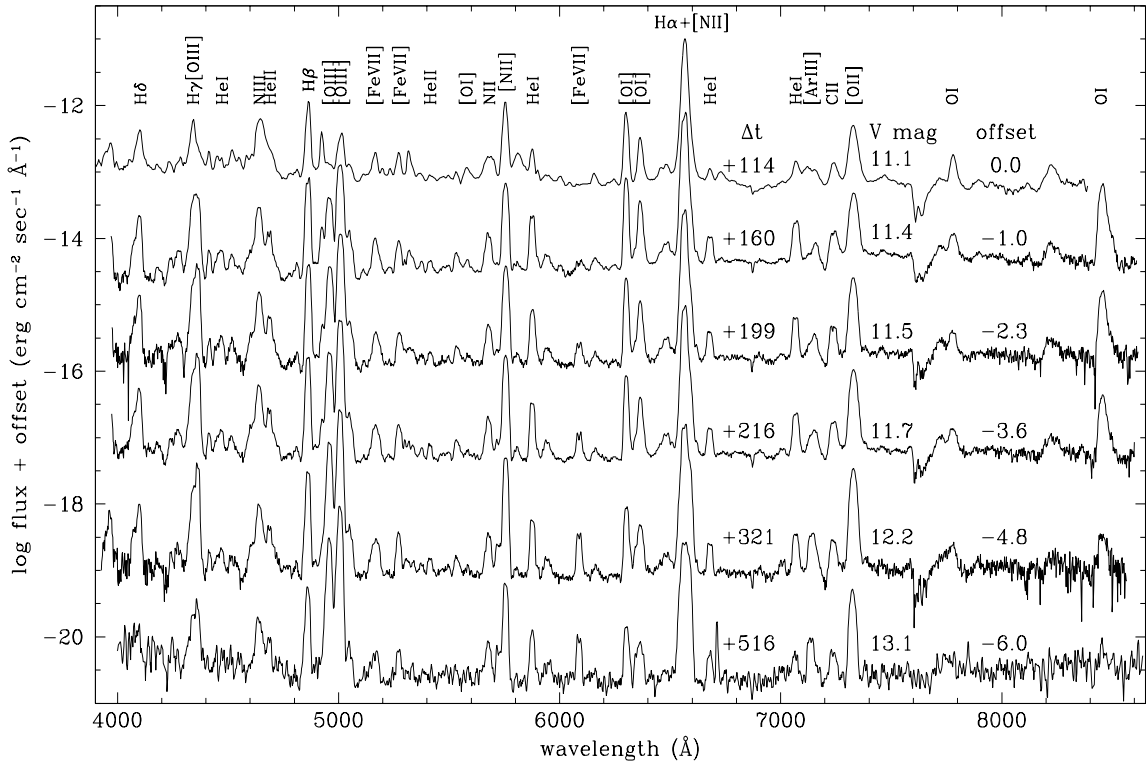}
\caption[]{Low-res spectroscopic evolution of V496 Sct during the optically thin
  branch (nebular condition) of the lightcurve, with time from optical maximum,
  the $V$ band magnitude and the offset in log flux are given for each spectrum.
  The major emission lines are identified.}
  \label{fig7}
  \end{figure*}

\subsubsection {Absorption systems during early decline}
The Echelle spectra offer the possibility to observe at high resolution the
evolution of the absorption components. As illustrated in Fig. 8,
when the nova reached its maximum brightness, the broad single absorption component of P Cygni line profile
was replaced by two components, whose intensity gradually faded in parallel
with the decline in brightness of the nova. A similar behaviour was exhibited by V2615 Oph where two absorption components are seen in addition to the emission component following the optical maximum (Munari et al. 2008). Fig. 8 presents line profiles for Fe II 5018 from the high resolution echelle spectra to illustrate the evolution of the two absorption components. Both absorption
components increased their negative radial velocity with time, with a linear
trend as illustrated in Fig. 9 and the best fit lines are given by
the following expressions:
\begin{equation}
V_A = -343 - (9.2 \times t)
\end{equation}
\begin{equation}
V_B = -763 - (18.5 \times t)
\end{equation}
where $t$ is the time after optical maximum.
The approaching conjunction with the Sun prevented further observations of
V496 Sct after our 2009 December 10, 22 days past optical
maximum, and the next set of observations were resumed from 2010 March 13. Novae usually display
different absorption systems, which behave similarly from object to object,
and that have been studied in detail by McLaughlin (1960, hereafter McL60),
who introduced a handy nomenclature for them.  An impressive graphical
representation of them has been offered by Hack and Struve (1970, Fig. 4i) from very high resolution observation of Nova Del 1967 by Ch. Fehrenbach. The A and B components of Eq. (1) and (2) shown in Fig. 9 nicely correspond to the principal and diffuse enhanced absorption
systems described by McL60, who noted a clear correlation between the
$t_2$, $t_3$ decline times and the mean velocity of these absorption systems.
McL60 also noted how the radial velocity of these absorption systems
generally increase with time, as seen here in V496 Sct.

The McL60 velocity relation for the principal system is $\log v_{\rm prin}$ =
3.57 - 0.5 $\log t_2$, and predicts a mean -485 km s$^{-1}$ for the $t_2$=59
day of V496 Sct. The agreement with Fig. 9 is evident, considering
in particular that the approaching conjunction with the Sun prevented to
extend the observations to later epoch characterized by larger radial
velocities for both systems. The McL60 velocity relation for the diffuse
enhanced system is $\log v_{\rm diff-enh}$ = 3.71 - 0.4 $\log t_2$, and
it predicts -1005 km s$^{-1}$ for the $t_2$=59 day of V496 Sct, again in good
agreement with Fig. 9.
Munari et al. (2008) have pointed out similar agreement of the observed velocities for the absorption systems of H$\alpha$ in case of V2615 Oph with the predicted values using the statistical relations by McL60.

For a few days around optical maximum, the high resolution spectra displayed
a rich ensemble of very sharp absorption lines of modest radial velocity
displacement, due to low ionization metals like Ti II, which will be
investigated elsewhere. They are similar to the transient heavy element absorption systems resulting from the episodic mass ejection from the secondary star seen in novae by Williams et al. (2008).

The permitted lines of Fe II are the strongest non-Balmer lines both during the pre-maximum rise, near optical maximum and early decline indicating $P_{fe}$ spectral class for V496 Sct during these phases (Williams et al. 1991; Williams, Phillips \& Hamuy 1994).

\subsubsection {Nebular phase}
The spectral evolution during the optically thin branch (nebular phase) is shown in Fig. 7. The evolution has been pretty standard, with
[O III] 4363, 4959, 5007, [N II] 5755, 6548, 6584, [O II] 7325 and [O I]
6300, 6364 being the dominant lines. The [O I] 6300/6364 flux ratio, that
during the early phases was close to 1 and indicative of a large optical
depth in the lines, with the thinning of the ejecta, the opacity increased toward the
3.1 normal ratio. The Ly $\beta$ fluorescent O I 8446 line, has remained
strong throughout the outburst and begun declining around day +200
when the optical thinning of ejecta reduced the trapping of Ly $\beta$
photons and therefore the fluorescent pumping of O I atoms. The ionization
conditions have been steadily increasing with advancing decline, with He I, He II
and [Fe VII] lines monotonically increasing in intensity with respect to
the other lines. The presence of a feable [Fe X] 6375 component could
compatible with the profile for the [O I] 6360 + [S II] 6347, 6371 blend
of the day +516 spectrum in Fig. 7, a firmer conclusion requiring
a spectrum of higher S/N and resolution.
The spectra shown in Fig. 7 show how the transition of V496 Sct from \emph{permitted} to \emph{nebular} phase occurred at an intermediate time after the last permitted spectrum of 2010 March 13 and the first nebular spectrum of 2010 April 28 where [O III] doublet lines $\lambda$4959 and $\lambda$5007 are seen prominently. We assign $A_o$ spectral class for V496 Sct (Williams et al. 1991; Williams, Phillips \& Hamuy 1994).

The presence of an emission feature at $\lambda$4924 coinciding with Fe II(42) line in the spectra taken during 2010 March till June is little puzzling. As the other prominent Fe II lines at $\lambda$4584 and $\lambda$5018 are absent as expected in the nebular phase, this feature is unlikely to be associated with Fe II multiplet. We would like to point out the presence of the emission feature at $\lambda$4924 as an unidentified feature similar to several such features seen in the spectra of many novae.


 \begin{figure}
\includegraphics[width=2.5in,height=2.5in, clip]{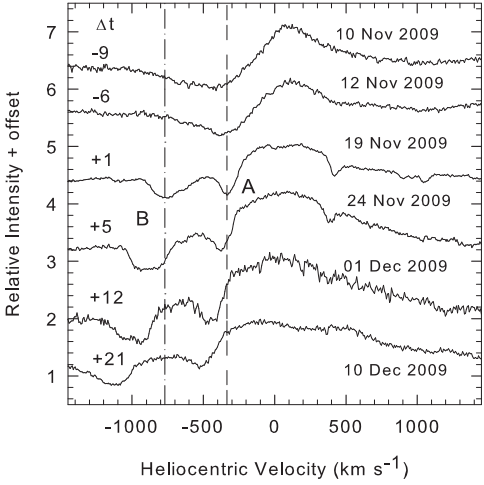}
  \caption[]{ Profiles of Fe II $\lambda$5018 line in the pre-maximum rise, near optical maximum and the early decline phases. The initial positions of the two narrow absorption components are shown by the broken lines. The time from optical maximum are given for each spectrum.}
  \label{fig7}
  \end{figure}

\begin{figure}
\includegraphics[width=2.5in,height=2.5in, clip]{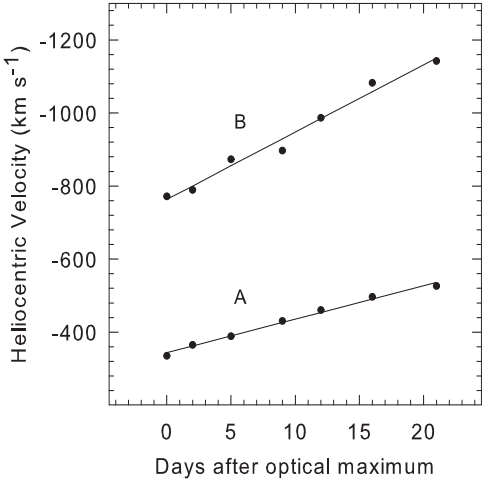}
  \caption[]{ The time evolution of radial velocities of the two absorption components of Fe II $\lambda$5018 line.}
  \label{fig7}
  \end{figure}

\subsection{Physical parameters}
The emission line fluxes of hydrogen and other elements can be used to estimate the physical parameters of the nova ejecta. In the early decline phase when the electron number densities are large, it is necessary to take in to account the optical depth $\tau$ while deriving the physical parameters. We determine the optical depth for [O I] $\lambda$6300 line using the formulation of Williams (1994), viz.,
\begin{equation}
\frac{j_{6300}}{j_{6364}} = \frac{1-e^{-\tau}}{1-e^{-\tau/3}}
\end{equation}
where j is the line emissivity. For the period 2010 April 28 to 2010 October 5
we get $\tau$ in the range 0.54-3.21. Now from the optical depth and the electron temperature we can estimate the mass of oxygen in the ejecta using the $\lambda$6300 line.
\begin{equation}
M_{O I} = 152d^{2} e^{22850/T_{e}} \times 10^{1.05E(B-V)}\frac{\tau}{1 - e^{-\tau}} F M_\odot
\end{equation}

where F is the flux of $\lambda$6300 line. Taking typical value of T$_{e}$ = 5000 K for the electron temperature (Ederoclite et al. 2006) we find M$_{O I}$ in the range 1.18$\times$10$^{-5}$-2.28$\times$10$^{-6}$ M$_\odot$.
The electron number density N$_{e}$ can be determined by [O III] line as given in Osterbrock (1989)
\begin{equation}
\frac{j_{4959}+j_{5007}}{j_{4363}} = 7.73 \frac{e^{3.29\times10^{-4}/T_{e}}}{1+4.5\times10^{-4}\frac{N_{e}}{T_{e}^{1/2}}}
\end{equation}
The values we obtained are in the range 10$^{4}$ to 10$^{6}$ cm$^{-3}$ close to the lower limit of the critical densities to give rise to nebular and auroral lines. This indicates that these lines are arising in relatively low density regions.
Following Osterbrock (1989) we have a relation between the intensity of the H$\beta$ emission line and the mass of hydrogen in the emitting nebula of pure hydrogen as
\begin{equation}
m(H)/M_\odot = \frac{d^2\times2.455\times10^{-2}}{\alpha^{eff}N_e}I(H\beta)
\end{equation}
where $\alpha$$^{eff}$ is the effective recombination coefficient and I(H$\beta$) is the flux for H$\beta$ line.
The mass of hydrogen m($H$) in the ejecta is (6.3 $\pm$ 0.2)$\times$10$^{-5}$M$_\odot$.
 As noted earlier in section 3.2 V496 Sct formed dust. The infrared observations by Russell et al. (2010) showed the presence of dust on 2010 February 10 which is still present on 2010 April 11 as indicated by the large (J-K) colour and the rising continuum. A sharp decline around 2010 February 8 seen in the V band light curve presented in Fig. 1 also indicates the onset of dust formation. It would be interesting to make an estimate of the dust mass $M_{dust}$ in V496 Sct and compare it with other novae that formed dust in their ejecta using the thermal component of the spectral energy distribution (SED). We adopt the method described by Woodward et al. (1993) that uses $(\lambda F_{\lambda})_{max}$ and $T_{dust}$ values obtained from the thermal component of the spectral energy distribution (SED). It is pertinent to point out that the present JHK photometric observations cover mostly the increasing part of the SED and thus the estimate of the temperature for the dust $T_{dust}$ likely to have large uncertainty. We obtain M$_{dust}$ $=$ 1-5$\times$10$^{-10}$M$_\odot$ for 2010 April 30 from the best fit value T$_{dust}$ $=$ 1500 $\pm$ 200 K (with $\chi$$^2$ minimization) for the temperature of the dust shell, $(\lambda F_{\lambda})_{max}$ = 2.62 $\times$10$^{-16}$ W $cm^{-2}$ and d = 2.9 kpc.
The estimated masses for different constituents of the ejecta like hydrogen, oxygen and dust derived from the optical and the infrared observations may be usefully utilized to derive the gas to dust mass ratio in novae. Gehrz et al. (1998) have presented a compilation of M$_{gas}$ and M$_{dust}$ along with ratio M$_{gas}$/M$_{dust}$ ranging from 5 in case of V705 Cas and 3$\times$10$^{4}$ in case of QU Vul. In case of V2362 Cyg, a very fast Fe II nova, Munari et al. (2008) have derived a value of 3$\times$10$^{5}$ for M$_{gas}$/M$_{dust}$. Taking the average fractional yield (by mass) of hydrogen to be 0.32 $\pm$ 0.10 for white dwarf masses ranging from 0.6 to 1.25 M$_\odot$ as per calculations of Jose \& Hernanz (1998) and Starrfield et al. (1997), the total gas mass based on the mass of hydrogen gas (determined here as 6.3$\times$10$^{-5}$M$_\odot$) is estimated to be 2.0 $\pm$ 0.6 $\times$10$^{-4}$M$_\odot$. Hence the gas to dust ratio is found to be M$_{gas}$/M$_{dust}$ $\sim$ 1.3-6.3 $\times$10$^{5}$ indicating that a small amount of dust was formed in V496 Sct comparable to 3 $\times$10$^{5}$ observed in the case of V2362 Cyg by Munari et al. (2008).

\begin{table*}
\caption[]{Fluxes of prominent emission lines relative to H$\beta$ = 100. The fluxes of the emission lines including H$\beta$ are corrected for E(B-V) = 0.57}
\begin{tabular}{cccccc}
\hline\\
Wavelength & Species  & 2009 Nov. 09 & 2009 Nov. 19  & 2009 Dec. 10 & 2010 Apr. 28  \\
(A) & &                 &  & &\\
\hline \\

3970   & Ca\,{\sc ii} and H$\epsilon$      & 28.3&39.4  & 30.0   &   \\
4101   & H$\delta$        & 45.2& 36.6&   7.2 & 54.2\\
4173   & Fe\,{\sc ii}(27) &27.7 &24.7& 18.0&  \\
4340   & H$\gamma$  &21.1 & 11.8&  36.2&141.5\\
4584   & Fe\,{\sc ii}(38) &48.2 &37.3        &   19.8 & \\
4635   & N\,{\sc iii}        &13.3 & 42.9&   14.2 &41.5\\
4686   & He\,{\sc ii}        &  & &&9.3\\
4861   & H$\beta$              &100   &100 & 100   & 100   \\
4924   & Fe\,{\sc ii}(42)             &31.3 &44.7&39.8 & 5.9       \\
4959   & [O\,{\sc iii}]              &   & &&    48.3 \\
5007   & [O\,{\sc iii}]             &     & & &     182.2\\
5018   & Fe\,{\sc ii}              &     36.8  &35.0   &58.5 & \\
5169+5176   & Fe\,{\sc ii} + Mg\,{\sc i} + N\,{\sc ii}          &36.8    &21.7  &64.2 &11.0    \\
5535   & Fe\,{\sc ii}(55) + N\,{\sc ii}        &3.6 &8.1&  6.1  &           \\
5577   &  O\,{\sc i}   &     &   &     3.7&\\
5675   & N\,{\sc ii}    &     &   &     &10.2\\
5755 & [N\,{\sc ii}](3)    &  & & &66.1\\
5876   & He\,{\sc i}              & &&  &22.0   \\
5890 & Na\,{\sc i}           &        &10.5 &11.4 &  \\
6157   & O\,{\sc i}       &        2.4  &10.7 &10.6 &\\
6243 & Fe\,{\sc ii} + N\,{\sc ii}   & 12.7 &29.0& 14.0&         \\
6300 & [O\,{\sc i}]    &           & &12.3&58.5\\
6363   		    & [O\,{\sc i}]      	&& & 6.8	&		28.0\\
6456              &    Fe\,{\sc ii}      & 6.0           & 19.9& 11.8 & \\
6563              & H$\alpha$       & 234.9   & 195.7& 240.3&  690.7 $^{a}$   \\
6678              & He\,{\sc i}      &   & &   &       8.5  \\
7065              &  He\,{\sc i}      &              & & &15.3\\
7330              & [O\,{\sc ii}]       & &  & & 44.9 \\
7774              & O\,{\sc i}       &   18.1  & 12.5  &19.1 & 5.1      \\
8446              & O\,{\sc i}       &    6.0  &14.2     & 16.9& 47.5  \\
8498              & Ca\,{\sc ii} triplet   &   & 20.3&5.4 &               \\
8542              & Ca\,{\sc ii} triplet       &              & 16.6& 9.1&\\

\hline
H$\alpha$/H$\beta$  &     &  2.4  &    2.0  & 2.4       & 6.9 \\
H$\beta$     & 10$^{-11}$erg cm$^{-2}$ s$^{-1}$        &16.6   &71.8 & 97.0   & 11.8   \\
\hline
\end{tabular}
\\
$^{a}$[NII] 6548, 6584 expected to contribute significantly to the overall flux of H$\alpha$ blend. \\
\end{table*}

\section{Summary}
We have presented near-infrared and optical spectroscopy and photometry of nova V496 Sct which erupted on 2009 November 8. From the optical lightcurve, the absolute magnitude and the distance to the nova are estimated to be $M_V = -7.0$ $\pm$ 0.2 and $d = 2.9$ $\pm$ 0.3 kpc respectively. The infrared and optical spectra indicate that the nova is of the Fe II class. Evidence is seen from the $JHK$ photometry for the formation of dust in the nova in 2010 April. In this context, the presence of emission lines from low ionization species like Na and Mg in the early spectra and subsequent formation of the dust supports the predictive property of these lines as indicators of dust formation as proposed by Das et al (2008). V496 Sct is one of the moderately fast Fe\,{\sc ii} class of novae ($t_2$ = 59 d) that showed CO emission before the dust formation.

The various phases of the spectral evolution of V496 Sct have been identified using the Tololo classification system for novae (Williams et al. 1991; Williams, Phillips \& Hamuy 1994). The permitted lines of Fe II were the strongest non-Balmer lines in the pre-maximum as well as the early decline phase indicating P$_{fe}$ class for the nova. The nova had evolved to the auroral phase A$_{o}$ in 2010 March as the [N II] 5755 auroral line was the strongest non-Balmer line. We note the absence of [Fe X] 6375 coronal emission line in the spectra taken as late as 2011 April 19. Thus the optical spectra show that the nova evolved in the P$_{fe}$A$_{o}$ spectral sequence.

\section*{Acknowledgments}

The research work at Physical Research Laboratory is funded by the Department of Space, Government of India. We would like to thank A. Frigo, V. Luppi, L. Buzzi, A. Milani, G. Cherini, A. Maitan, L. Baldinelli (ANS Collaboration). We thank the referee for the helpful
comments.

\label{lastpage}
\end{document}